\documentclass[vecphys]{svmult}

\usepackage{makeidx}         
\usepackage{graphicx}        
\usepackage{multicol}        
\usepackage{cite}            
\usepackage{bm}            
\usepackage[bottom]{footmisc}

\makeindex             

\begin{document}

\title{Quantized vortices and quantum turbulence}
\author{Makoto Tsubota$^1$ \and Kenichi Kasamatsu$^2$}
\institute{Department of Physcs, Osaka City University
\texttt{tsubota@sci.osaka-cu.ac.jp}
\and Department of physics, Kinki University 
\texttt{kenichi@phys.kindai.ac.jp}}

\maketitle

\begin{abstract}
We review recent important topics in quantized vortices and quantum turbulence in 
atomic Bose--Einstein condensates (BECs). They have previously been studied for a long time in superfluid helium. Quantum turbulence is currently one of the most important topics in low-temperature physics. Atomic BECs have two distinct advantages over liquid helium for investigating such topics: quantized vortices can be directly visualized and the interaction parameters can be controlled by the Feshbach resonance. A general introduction is followed by a description of the dynamics of quantized vortices, hydrodynamic instability, and quantum turbulence in atomic BECs.
\end{abstract}

\noindent

\section{Introduction}
\label{introduction}
Bose--Einstein condensation is often considered to be a macroscopic quantum phenomenon
because bosons occupy the same single-particle ground state below 
the critical temperature for Bose--Einstein condensation so that they have a macroscopic wave function (order parameter) $\Psi({\bm r},t)=|\Psi({\bm r},t)| e^{i \theta({\bm r},t)}$ that extends 
over the entire system. Here, the absolute squared amplitude $|\Psi|^2 = n$ gives the condensate 
density and the gradient of the phase $\theta({\bm r},t)$ gives 
the superfluid velocity field ${\bm v}_s=(\hbar/m) \nabla \theta$ with boson mass $m$ as the potential flow.
Since the macroscopic wave function should be single-valued for the space coordinate ${\bm r}$, 
the circulation $\Gamma = \oint {\bm v}_s \cdot d{\bm \ell}$ for an arbitrary closed loop in the fluid will be quantized with the quantum $\kappa=h/m$.
A vortex with such quantized circulation is known as a quantized vortex. 
Any rotational motion of a superfluid is sustained only by quantized vortices.
Hydrodynamics dominated by quantized vortices is called {\it quantum hydrodynamics} (QHD), and turbulence comprised of quantized vortices is known as {\it quantum turbulence} (QT). 

A quantized vortex is a stable topological defect that is a characteristic of a Bose--Einstein condensate (BEC).
It differs from a vortex in a classical viscous fluid in the following three ways. 
First, unlike a classical vortex that can have an arbitrary circulation, the circulation of a quantized vortex is quantized.
Second, since a quantized vortex is a vortex of inviscid superflow 
it cannot decay by viscous diffusion of vorticity, which occurs in classical fluids.
Third, the core of a quantized vortex is very thin, being of the order of the coherence length (i.e., only a few angstroms in superfluid $^4$He and submicrometer in atomic BECs).
Since the vortex core is very thin and does not decay by diffusion, the position of a quantized vortex in the fluid can always be identified.

Since any rotational motion of a superfluid is sustained by quantized vortices, QT usually takes the form of a disordered tangle of quantized vortices.
QT is currently the most important research topic in QHD, which is an area  in the field of low-temperature physics.
The turbulence in a classical fluid, which is known as classical turbulence (CT), has been extensively studied in a number of fields, but it is still not well understood \cite{Frisch}. 
This is mainly because turbulence is a complicated dynamical phenomenon that is highly nonlinear. 
Vortices may represent the key for understanding turbulence,
but they are not well defined for a classical viscous fluid.
They are unstable and appear and disappear repeatedly. The circulation is not conserved and varies between vortices. 
Comparison of QT and CT reveals definite differences, which demonstrates the importance of studying QT. 
QT consists of a tangle of quantized vortices that have the same conserved circulation. 
Thus, QT can be  easier to study than CT and it offers a much simpler model of turbulence than CT. 

Quantized vortices and QT have historically been studied in superfluid helium.
However, the realization of Bose--Einstein condensation in trapped atomic gases in 1995 provided another important system for studying quantized vortices and QT.
The existence of superfluidity has been confirmed by creating and observing quantized vortices in atomic BECs
and a lot of effort has been devoted to studying lots of fascinating problems.
Atomic BECs have several advantages over superfluid helium,
the most important being that modern optical techniques can be used to directly control their properties and to visualize quantized vortices.

In a weakly interacting Bose system at zero temperature, the macroscopic wave function 
$\Psi({\bm r},t)$ obeys the Gross--Pitaevskii (GP) equation \cite{Gross,Pitaevskii}: 
\begin{equation}
i \hbar \frac{\partial \Psi({\bm r},t)}{\partial t} = \left( - \frac{\hbar ^2}{2m}\nabla^2 
+ U ({\bm r})+g |\Psi({\bm r},t)|^{2}- \mu \right) \Psi({\bm r},t). \label{eq-gp}
\end{equation}
Here, $U ({\bm r})$ represents the external potential (trapping potential, obstacle potential, etc.), 
$g=4\pi \hbar^2 a/m$ denotes the strength 
of the interaction characterized by the s-wave scattering length $a $, 
and $\mu$ is the chemical potential. 
The only characteristic length scale of the GP model is the coherence length.
It is defined by $\xi=\hbar/(\sqrt{2mg}| \Psi |)$ and gives the vortex core size. 
The GP equation has been used to interpret many properties 
of BECs in a dilute atomic gas \cite{PethickSmith,PitaevskiiStringari}.
It can explain both vortex dynamics and vortex core
phenomena, such as reconnection and nucleation.

This chapter is organized as follows. Section \ref{dynamics} describes the dynamics of quantized vortices. 
It is very difficult to directly observe their dynamics in superfluid $^4$He and $^3$He, which
is one important advantage of atomic BECs over superfluid helium.
In Sec. \ref{instability}, we discuss hydrodynamic instability and QT in atomic BECs.
The last section presents the conclusions.

\section{Dynamics of quantized vortices}
\label{dynamics}
Since the realization of atomic BECs, many researchers have formed
vortices by stirring a condensate or phase engineering.
The details of these studies have been reviewed in other articles \cite{Fetter,KTreview}. 
This chapter describes some recent contributions to this topic.

\subsection{Dynamics of a single vortex and vortex dipoles}
To gain insight into diverse superfluid phenomena, 
it is essential to understand the dynamics of a single vortex or vortex dipoles (i.e., vortex--antivortex pairs). 
A few important observations have recently been made.
Neely {\it et al.} nucleated vortex dipoles in an oblate BEC by forcing superfluid flow around 
a repulsive Gaussian obstacle generated by a focused blue-detuned laser beam \cite{Neely}.
The nucleated vortex dipole propagates in a BEC cloud for many seconds; its continuous trajectory 
was found to be consistent with a numerical simulation of the GP model.
Freilich {\it et al.} observed the real-time dynamics 
of vortices by repeatedly extracting, expanding, and imaging small fractions 
of the condensate to visualize the motion of the vortex cores \cite{Freilich}. 
They nucleated vortices via the Kibble--Zurek mechanism \cite{Kibble, Zurek} in which rapid 
quenching of a cold thermal gas through the BEC phase transition causes topological defects to nucleate \cite{Weiler}.
This produces single-vortex precesses in the cloud, whose frequency is in 
good agreement with a simple theoretical analysis \cite{Fetter01}.
Freilich {\it et al.} also nucleated vortex dipoles and observed their real-time dynamics. 
This new technique enables quantitative comparisons to be performed between experiment 
and theory for vortex dynamics \cite{Kuopanportti,Middelkamp}.

\begin{figure}[ht]
\includegraphics[height=0.21\textheight]{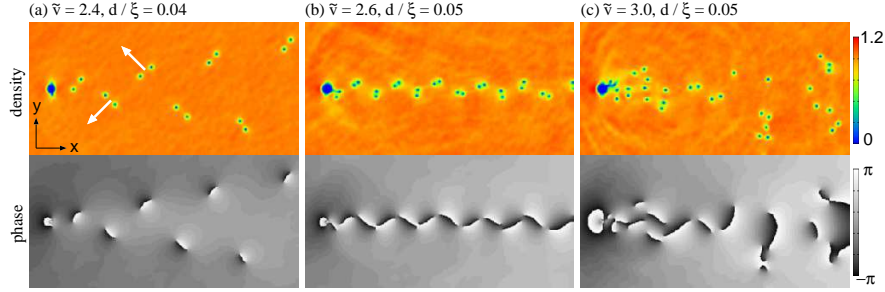}
\caption{Density and phase distributions of a condensate past an obstacle potential. The velocity and potential width are (a) $(\tilde{v}, d/\xi)=(2.4, 0.04)$, (b) $(2.6, 0.05)$, and (c) $(3.0, 0.05)$, where $\tilde{v}=v(10^3m/gn_0)^{1/2}$ and $\xi=\hbar (10^3/(mgn_0)^{1/2}$ with $n_0$ the atom density without perturbation. 
The white arrows in (a) indicate the directions in which the vortex pairs move. 
The field of view is $6\xi \times 3\xi$. [Sasaki, Suzuki and Saito: Phys. Rev. Lett. {\bf 104}, 150404 (2010), reproduced with permission.
Copyright 2010 by the American Physical Society].}
\label{Saito}
\end{figure}

Vortex dipole generation has been investigated theoretically and numerically in some recent studies based on the GP model. 
Sasaki {\it et al.} studied vortex shedding from an obstacle potential moving in a uniform BEC \cite{Sasaki10}. 
The flow around the obstacle is laminar when the velocity of the potential is sufficiently low.
When the velocity exceeds a critical velocity of the order of the sound velocity, the potential commences to emit vortex trains.
The manner in which vortices are emitted depends on the velocity and the width of the potential.
The first pattern is the emission of the alternately inclined vortex pairs in a V-shaped wake, as shown in Fig. \ref{Saito}(a).
The second pattern is produced by sequential shedding of two vortices having the same circulation. 
The two vortices rotate about their center without varying their separation, forming a train of vortices 
similar to a B\'{e}nard--von K\'{a}rm\'{a}n vortex street, as shown in Fig. \ref{Saito}(b).
For a wide potential with a high velocity, the periodicity disappears, as shown in Fig. \ref{Saito}(c).
Aioi {\it et al.} subsequently proposed controlled generation and manipulation of vortex dipoles by using several 
Gaussian beams from a red (attractive potential) or blue (repulsive potential) detuned laser \cite{Aioi11}.
For example, when a red-detuned beam moves through a BEC cloud above a critical velocity, a vortex dipole nucleates at the head of the potential. 
In contrast, a blue-detuned beam generates vortex dipoles on both sides of the potential. 
Double beams can generate various kinds of dipole wakes depending on the velocity and the distance of the two beams.
The trajectory of emitted vortices can be modified by another potential.

\subsection{Dynamics of vortices generated by an oscillating obstacle potential}
Recently, in the field of superfluid $^4$He and $^3$He, several groups have experimentally studied QT generated by oscillating structures such as wires, spheres, and grids \cite{SVreview}.
Despite significant  differences between the structures used, their responses with respect to the alternating drive have revealed some surprising common phenomena.
The response was laminar at low driving rates, whereas it became turbulent at high driving rates. 

\begin{figure}[ht]
\includegraphics[height=0.28\textheight]{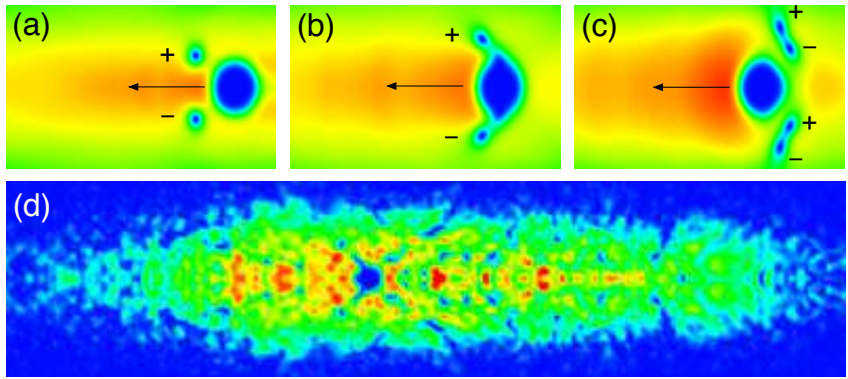}
\caption{Reconnection of vortex pairs about an oscillating potential. The symbols - and + indicate vortices with clockwise and counterclockwise circulations, respectively. The black arrows indicate the direction in which the potential moves. 
The horizontal and vertical dimensions of (a), (b) and (c) are 27.0 and 14.0 micrometer, while those of (d) are 145 and 34.0 micrometer. 
When the potential moves to the right, it generates a vortex pair with an impulse in the same direction as the potential. 
When the potential (a) ($t$=31.4 ms) changes direction and moves to the left, (b) ($t$=38.3ms) 
it collides with the pair and (c) ($t$=43.2 ms) emits another pair. 
These two pairs reconnect to form new pairs that move away from the potential. 
(d) The BEC cloud eventually becomes filled with such vortices. 
[Fujimoto and Tsubota: Phys. Rev. A {\bf 83}, 053609 (2011), reproduced with permission.
Copyright 2011 by the American Physical Society].} 
\label{Fujimoto}
\end{figure}

This strategy can also be applied to trapped atomic BECs.
Fujimoto and Tsubota numerically investigated the two-dimensional dynamics of trapped BECs 
induced by an oscillating repulsive Gaussian potential \cite{Fujimoto10, Fujimoto11} and
found a strong dependence on the amplitude and frequency of the potential.
Unlike a potential with constant velocity \cite{Sasaki10}, an oscillating potential continually 
sheds vortex pairs with alternating impulses, a typical example of which is shown in Fig. \ref{Fujimoto}.
The nucleated pairs form new vortex pairs through reconnections that move away 
from the obstacle potential and toward the condensate surface.
The BEC cloud eventually becomes full with such vortices, as shown in Fig. \ref{Fujimoto}(d). 
An oscillating potential is thus a useful tool for generating QT in a trapped BEC.
 
\subsection{Kelvin wave dynamics}
Kelvin waves are three-dimensional excitations along a vortex line. 
Kelvin waves play an important role in dissipation in superfluid 
helium at very low temperatures in which the normal fluid component is negligible \cite{TKreview}.
Kelvin waves have been experimentally demonstrated in a trapped BEC \cite{Bretin03} by
exciting a collective quadrupole mode to a single-vortex state that subsequently decays 
to Kelvin modes by a nonlinear Beliaev process. This was supported by numerical analysis based on the
Bogoliubov--de Gennes equation \cite{Mizushima03} and the GP equation \cite{Simula}.
Rooney {\it et al.} recently studied the dynamics of Kelvin waves using the stochastic projected GP equation \cite{Rooney11}.
They showed that Kelvin waves can be suppressed by tightening the confinement of the trap along the vortex line, 
which drastically reduces the vortex decay rate as the system becomes two-dimensional.
This behavior is consistent with observations of the decay of vortex dipoles \cite{Neely}.

\section{Hydrodynamic instability and quantum turbulence}
\label{instability}
Some theoretical ideas for achieving QT in trapped BECs have been proposed. 
The turbulent state may be generated during the formation 
of a condensate from a nonequilibrium non-condensed Bose gas by 
rapid quenching \cite{Berloff}. 
Conversely, turbulence is expected to be generated during 
the {\it destruction} process from the equilibrium condensed state. 
This subsection describes recent studies on hydrodynamic instability and the resulting nonlinear dynamics 
that may lead to QT. These instabilities are formed by applying an external driving force to 
condensates or by complicated interactive phenomena between multicomponent BECs
(some of which are analogs of well-known phenomena in classical hydrodynamics).

\subsection{Methods to produce turbulence in trapped BECs}
A major problem when investigating QT in atomic BECs is the difficulty in applying a dc velocity field in superfluid helium.
Here, we summarize some proposals for and a recent experimental realization 
of QT generation in a trapped single-component BEC. 
\paragraph{Simple rotation}

An important way for generating vortices in trapped BECs is to rotate the external potential \cite{Madison,Abo,Hodby}.
However, rotation alone cannot lead to QT because it generates an ordered vortex lattice along the rotational axis \cite{Tsubota}, 
which is the equilibrium state in the corresponding rotating frame. 
Nevertheless, Parker and Adams suggested the emergence and decay of turbulence in a 
BEC under a simple rotation, starting from a vortex-free equilibrium BEC \cite{Parker}.
A numerical simulation based on the energy-conserving GP equation suggests the existence of a turbulent regime that contains many vortices and high-energy-density fluctuations (sound field) on a route to the ordered vortex lattice.

\paragraph{Two-axis rotation}

\begin{figure}[ht]
\includegraphics[height=0.3\textheight]{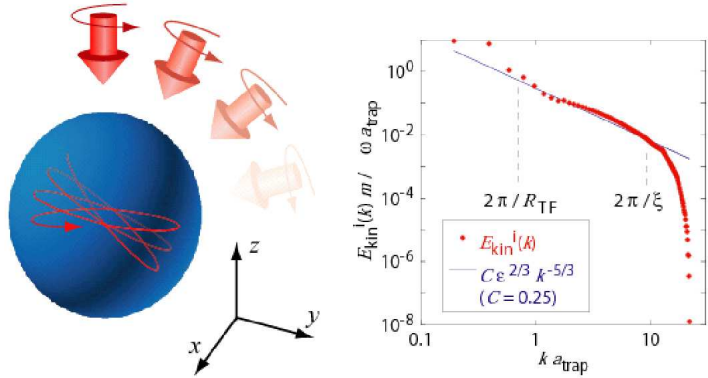}
\caption{QT in atomic BECs. 
(a) A method for realizing steady turbulence in an atomic BEC subject to precession. 
A BEC is trapped in a weakly elliptical harmonic potential. 
A rotation is applied along the $z$-axis followed by a rotation along the $x$-axis. 
(b) Energy spectrum of steady QT obtained by two-axis rotation. 
The dots represent the numerically obtained spectrum for an incompressible kinetic energy, 
while the solid line is the Kolmogorov spectrum. Here, $R_{\rm TF}$ is the 
Thomas--Fermi radius and $a_{\rm trap}=\sqrt{\hbar/m\omega}$ is the characteristic length scale of the trap. 
}
\label{twoaxispros}
\end{figure}

Since the above turbulence is generated during the ordering process, it is not steady turbulence. 
Kobayashi and Tsubota suggested performing rotations about two axes \cite{Kobayashi2007}, 
as shown in Fig. \ref{twoaxispros}(a). When the spinning and precessing 
rotational axes are perpendicular, the two rotations do not commute and thus cannot be represented by their sum. 
This situation can be modeled by simulations based on the GP equation:
\begin{equation}
i \hbar \frac{\partial \Psi}{\partial t} = \left( - \frac{\hbar \nabla^2}{2m} + U 
+g |\Psi|^2 - {\bm \Omega} \cdot {\bm L} \right) \Psi, \label{GPErot}
\end{equation}
where $ {\bm L}$ is the angular momentum and the rotation vector is written as
${\bm \Omega}(t)=(\Omega_x,\Omega_z\sin\Omega_xt,\Omega_z\cos\Omega_xt)$
with frequencies of $\Omega_z$ and $\Omega_x$ for the first and second rotations, 
respectively. The inclusion of these non-commuting rotations 
and phenomenological dissipation which is effective only at scales smaller than the healing length 
successfully generate steady turbulence \cite{Kobayashi2007}. 

\paragraph{Donnelly--Glaberson instability}

When both rotation and a linear velocity are applied to the system,
the right-hand side of the GP equation Eq. (\ref{GPErot}) will have the term $- {\bm V} \cdot {\bm p}$ 
in the corresponding {\it helically} moving frame.
For ${\bm V} = V \hat{\bm z}$ 
the dispersion of the vortex waves (Kelvin waves) on a straight vortex line parallel to 
the $z$-axis behaves as $\omega - V k_{z}$ and its frequency 
will become negative above a certain critical velocity. 
This is known as the Donnelly--Glaberson instability in superfluid helium
and it amplifies Kelvin waves \cite{Tsubota03}. 
If the rotating BEC contains a vortex lattice, the amplified Kelvin wave induces 
reconnections of adjoining vortex lines, eventually leading to a turbulent state \cite{Takeuchi09}. 

\paragraph{Combined rotation and oscillating excitation}

As a method that enables better control of turbulence, Henn {\it et al.} introduced an external 
oscillatory potential to an $^{87}$Rb BEC \cite{Henn2009a,Henn2009b}.
This oscillatory field induced a successive coherent mode excitation in a BEC.
They observed that increasing the amplitude of the oscillating field and the excitation period
increased the number of vortices and eventually lead to the turbulent state \cite{Henn2009b}.
In the turbulent regime, they observed a rapid increase in the number of vortices followed 
by proliferation of vortex lines in all directions, 
where many vortices with no preferred orientation formed a vortex tangle. 
The oscillatory excitation (which mainly consists of oscillation, rotation,
and deformation) nucleated vortices. However, it is still not known theoretically how
the turbulence is generated.


\subsection{Signature of quantum turbulence}

It is important to determine whether the highly excited state is really QT.
Several methods are used to identify the turbulent state; they are discussed in this subsection.

In most numerical simulations, apart from observing a random 
configuration of vortices, the turbulent regime has been identified by checking 
if its incompressible kinetic energy spectrum obeys
Kolmogorov's $-5/3$ law \cite{Frisch}. When the condensate 
wave function is written in the form $\Psi({\bm r}, t) = \sqrt{ n({\bm r},t)} e^{i \theta({\bm r}, t)}$, 
the kinetic energy is expressed by the sum $E_{\rm kin}=E_{k}+E_{q}$, 
where $E_k= (\hbar^2/2m) \int d{\bm r} |\sqrt{n} \nabla \theta |^2$ 
denotes the superfluid kinetic energy and $E_q =  (\hbar^2/2m) \int d{\bm r} |\nabla \sqrt{n} |^2$ is
the quantum pressure energy.
The vector field $\sqrt{n} \nabla \theta$ can be divided into incompressible (solenoidal) 
and compressible (irrotational) components:
$\sqrt{n} \nabla \theta = (\sqrt{n} \nabla \theta)^{i} + (\sqrt{n} \nabla \theta)^{c}$, 
where $\nabla \cdot (\sqrt{n} \nabla \theta)^{i} =0$ and $\nabla \times (\sqrt{n} \nabla \theta)^{c} =0$. 
Thus, the incompressible and compressible kinetic energies are defined by
$E_k^{i,c} =  (\hbar^2/2m) \int d {\bm r} |  (\sqrt{n} \nabla \theta)^{i,c}  |^2$.
They correspond to the kinetic energies in the
vortices and the sound waves, respectively. Since the compressible
and incompressible fields are mutually orthogonal, it follows
that $E_k=E_k^i +E_k^c$. The kinetic energy spectrum as a function
of wave number $k$ is defined by
\begin{equation}
\epsilon_k^{i,c} (k) = \frac{\hbar^2}{2m} \int k^2 \sin \theta d\theta d\phi \left| \int 
e^{i {\bm k} \cdot {\bm r}}  (\sqrt{n} \nabla \theta)^{i,c} \frac{d{\bm r}}{(2\pi)^3} \right|,
\end{equation}
such that $E_{k}^{i,c} = \int_{0}^{\infty} \epsilon_{k}^{i,c}(k) dk$. 
The Kolmogorov law states that the incompressible energy spectrum obeys 
the power law $ \epsilon_{k}^{i}(k) \sim k^{\nu}$ where $\nu=-5/3$ over the inertial range of $k$. 
For a trapped BEC, the inertial range that follows the Kolmogorov law 
is determined by the Thomas--Fermi radius $R_{\rm TF}$ 
and the coherence length $\xi=\hbar/2m\mu$ \cite{Kobayashi2007} [see the right panel of Fig. \ref{twoaxispros}]. 

The structure of QT is reflected in the time dependence of the decay 
of the total vortex line density $L$ after turning off the excitation that
sustains the turbulence. Correlations of vortex tangles can be classified 
into two kinds \cite{Vinen}: correlated and uncorrelated tangles.  In a correlated tangle,
turbulent energy is concentrated in the ``classical" length scale range that is larger than the 
intervortex distance $l$, where the correlated tangle  exhibits a Kolmogorov spectrum. 
The energy is then transferred to smaller scales by a Richardson
cascade and $L$ decays as $t^{-3/2}$. 
In an uncorrelated tangle, the turbulent energy is associated with a
random vortex tangle with spacing $l$ and it is concentrated in the ``quantum" range 
of length scales smaller than $l$. Then, $L$ decays as $t^{-1}$. 
Thus, observing the scaling behavior of $L$ on $t$ provides useful information 
about QT.

In addition, White {\it et al.} \cite{White} showed that QT is characterized by a power-law behavior of the 
probability density function of the velocity field ${\bm v}_s({\bm r}) =(\hbar/m)  \nabla \theta ({\bm r})$, 
whereas classical turbulence obeys Gaussian velocity statistics. 
This non-Gaussian behavior originates from the singular nature of a quantized vorticity 
with a $1/r$ velocity field, which appears in the high-velocity region determined by 
the intervortex distance $v \sim \kappa/\pi l$ \cite{Adachi}. 

It is difficult to measure the above statistical or scaling properties experimentally. 
Henn {\it et al.} observed another remarkable feature of turbulent condensates: 
suppression of aspect ratio inversion during free expansion after turning 
the trapping potential off \cite{Henn2009b}. 
Despite the asymmetric expansion (from a cigar shape to a pancake shape) 
of a conventional BEC or isotropic 
expansion of a thermal cloud, the turbulent state exhibited a self-similar expansion 
that preserved the initial aspect ratio. 
Although a quantitative theoretical understanding of this effect has yet to be fully realized \cite{Caracanhas}, 
it represents a remarkable new effect in the turbulent regime.

\subsection{Hydrodynamic instability in multicomponent BECs}

Multicomponent atomic BECs can be created in cold-atom systems with, for example, 
multiple hyperfine spin states or a mixture of different atomic species. 
Such systems yield a rich variety of superfluid dynamics due to the intercomponent interaction. 
Two-component BECs are the simplest multicomponent system. 
Schweikhard {\it et al.} experimentally investigated the vortex-lattice dynamics of two interacting 
and rotating condensates by transferring some of the initial population of $^{87}$Rb BECs
with a vortex lattice to its other hyperfine state via a coupling pulse \cite{Schweikhard}. 
They observed the ordering dynamics change from a triangular lattice structure to a 
stable square lattice through the transient turbulent regime. 

Recent experimental advances have provided more controllable ways to 
study the rich dynamics of two-component BECs. 
External potentials can be prepared that can act independently on both components;
this enables initial conditions to be prepared
that are suitable for studying a particular problem. 
In addition, the intra- and inter-component interactions can be tuned with the help of 
the Feshbach resonance \cite{Thalhammer,Papp,Tojo}. 
This allows phase separation to be performed 
and interface phenomena 
between two superfluids to be studied in a well-controlled manner. 

In the mean-field theory, a two-component BEC is described by 
macroscopic wave functions $\Psi_i$, 
where the subscript $i$ refers to each component ($i=1,2$).
The Lagrangian for this system is given by
\begin{equation} \label{Lag}
L = \int d\bm{r} \left( P_1 + P_2 - g_{12} |\Psi_1|^2 |\Psi_2|^2
\right),
\end{equation}
where
\begin{equation}
P_i = i \hbar \Psi_i^* \frac{\partial \Psi_i}{\partial t} +
\frac{\hbar^2}{2 m_i} \Psi_i^* \nabla^2 \Psi_i - U_i |\Psi_i|^2 -
\frac{g_{ii}}{2} |\Psi_i|^4
\end{equation}
with $m_i$ and $U_i$ being the atomic mass and the external potential
of the $i$th component, respectively.
The intra- and inter-component interaction parameters have the form
\begin{equation}
g_{ij} = 2\pi \hbar^2 a_{ij} (m_i^{-1} + m_{j}^{-1}),
\end{equation}
where $a_{ij}$ is the s-wave scattering length between the $i$th and
$j$th components; we assume $a_{ij} > 0$ in the following. 
For homogeneous condensates, the condensates are miscible and immiscible when $g_{11}g_{22} > g_{12}^2$ 
and $g_{11}g_{22} < g_{12}^2$, respectively. 
Below, we review the characteristic hydrodynamic instability 
that occurs in each condition. 

\subsubsection{Counter-superflow instability}
When two-component BECs coexist with a relative velocity, they 
exhibit dynamic instability above a critical relative velocity \cite{Law}. 
This phenomenon is known as a counter-{\it superflow} instability (CSI). 
Takeuchi {\it et al.} suggested that the nonlinear dynamics triggered by the CSI 
generates a binary QT composed of coreless vortices and that thus has a continuous velocity 
field \cite{TakeuchiCSI,Ishino}. Hamner {\it et al.} realized the CSI experimentally for quasi-1D geometry and observed the generation
of shock waves and dark--bright solitons \cite{Hamner}. 

The CSI can be understood from the Bogoliubov spectrum for a system of a
uniform two-component BEC with a relative velocity \cite{Law}.
The functional derivative of $\int dt L$, where $L$ is the Lagrangian in
Eq. (\ref{Lag}), with respect to $\Psi_i^*$ gives the GP equation, 
\begin{eqnarray}
i \hbar \frac{\partial \Psi_i}{\partial t}  = \left(-\frac{\hbar^2}{2m_i}{\bm \nabla}^2+\sum_{j=1,2} g_{ij}|\Psi _j|^2\right)\Psi_i.
\label{eq:GP}
\end{eqnarray}
In this subsection, we assume $U_i = 0$ and the miscible condition $g_{11} g_{22} > g_{12}^2$. 
The wave functions $\Psi_i=\Psi_{i0}$ in a stationary state can be written as
\begin{eqnarray}
\Psi_{i0}=\sqrt{n_{i0}}e^{i\left(m_{i}{\bm V}_{i}\cdot{\bm r}-\mu_i t \right)/\hbar}
\label{eq:steady}
\end{eqnarray}
with the velocity ${\bm V}_i$ and the chemical potential $\mu_i$ of the $i$th component. 
Counter superflow occurs when ${\bm V}_1 \neq {\bm V}_2$. 
We consider a small excitation above the stationary state as $\Psi _{i}=\Psi_{i0}+\delta\Psi _{i}$,
 where we write the excitation of the wave functions $\delta\Psi _{i}$ in the form
 \begin{eqnarray}
 \delta \Psi_i =e^{i(m_{i}{\bm V}_{i}\cdot {\bm r}-\mu _{i} t)/\hbar}
\left[ u_{i}e^{i({\bm k}\cdot{\bm r}-\omega t)}-v_{i}^{\ast}e^{-i({\bm k}\cdot{\bm r}-\omega t)} \right].
 \label{eq:delta}
 \end{eqnarray}
By linearizing the GP equation (\ref{eq:GP}) with respect to $\delta\Psi_{i}$, 
we obtain the Bogoliubov-de Genne equations:
\begin{eqnarray}
\left( \frac{\hbar^2 k^2}{2m_i} + \hbar \bm{k} \cdot \bm{V}_i 
+ g_{ii} n_{i0} \right) u_{i} - g_{ii} n_{i0} v_{i} 
+ g_{ij} \sqrt{n_{i0} n_{j0}} \left( u_{j} - v_{j} \right)  =  
\hbar \omega u_{i},  \label{bogo2-1} \\
\left( \frac{\hbar^2 k^2}{2m_i} - \hbar \bm{k} \cdot \bm{V}_i 
+ g_{ii} n_{i0} \right) v_{i} - g_{ii} n_{i0} u_{i} 
- g_{ij} \sqrt{n_{i0} n_{j0}} \left( u_{j} - v_{j} \right) = 
-\hbar \omega v_{i}. \label{bogo2-2}
\end{eqnarray}
Diagonalizing the eigenvalue equations (\ref{bogo2-1}) and (\ref{bogo2-2}), 
we obtain the Bogoliubov excitation spectrum.
For simplicity, we set $m_1 = m_2 \equiv m$, 
$g_{11}=g_{22} \equiv g$, and $n_{10} = n_{20} \equiv n_{0}$, and 
neglect the center-of-mass velocity of the two condensates. 
The eigenvalue of Eqs. (\ref{bogo2-1}) and (\ref{bogo2-2}) 
then has the simple form:
\begin{eqnarray}
\hbar^2\omega^2=\epsilon^2+\frac{1}{4} \hbar^2 k_{||}^2V_{R}^2 \pm 
\sqrt{\epsilon^2\hbar^2 k_{||}^2V_{R}^2+4g_{12}^2n_0^2\epsilon_0^2},
\label{eq:dispersion}
\end{eqnarray}
where ${\bm V}_{R}={\bm V}_{1} - {\bm V}_{2}$ is the relative velocity
and we denote $\epsilon^2=\epsilon_0(\epsilon_0+2gn_0)$, 
$\epsilon_0= \hbar^2k^2/ 2m$, and $k^2=k_{||}^2+k_{\bot}^2$ 
with $k_{||}=({\bm k}\cdot {\bm V}_{R})/V_{R}$ and $k_{\bot} \geq 0$. 
The system is dynamically unstable when the excitation frequency 
$\omega$ becomes imaginary (i.e., $\omega^2<0$), which reduces to 
$\epsilon_- < \hbar k_{||} V_{R}/2 < \epsilon_+$ with 
$\epsilon_{\pm}=\sqrt{\epsilon_0[\epsilon_0+2(g\pm g_{12})n_0)]}$.
This inequality determines the unstable region 
in wavenumber space $(k_{||}, k_{\bot})$, where $V_{R}$ exceeds 
the critical relative velocity $V_c=V_-$ with $V_{\pm}=2\sqrt{gn_0/m}\sqrt{1\pm g_{12}/g}$. 
The total momentum density $\delta {\bm J}$ carried by the excitation can be 
defined as $\delta {\bm J}=\delta {\bm J}_1+\delta {\bm J}_2$ with a change 
in the momentum density $\delta {\bm J}_j \equiv \hbar {\bm k}(|u_i|^2-|v_i|^2)$ of the $i$th component.
The unstable modes, which trigger CSI, 
should satisfy the condition $\delta {\bm J}=0$ with $\delta {\bm J}_1=-\delta {\bm J}_2\neq 0$ 
due to the law of momentum conservation. Since the CSI 
is the dynamic instability triggered by unstable modes 
with an imaginary part ${\rm Im}~\omega \neq 0$ \cite{Law}, the amplification of the unstable modes 
will exponentially enhance the momentum exchange.

\begin{figure}[ht]
\includegraphics[height=0.11\textheight]{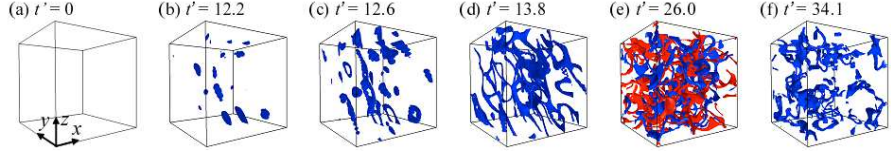}
\caption{Characteristic nonlinear dynamics 
of CSI of two-component BECs.
The parameters are set to $m_1=m_2=m$, 
$g_{11}=g_{22}=g$, and $g_{12}=0.9g$, similar 
to those in experiments. In the numerical simulation, the initial state 
is prepared by adding small random noise to the stationary wave function 
$\Psi_{i0}$ with $n_{10}=n_{20}=n_{0}$ and ${\bm V}_1=-{\bm V}_2$.
The panels shows the time development of low-density isosurfaces 
with $n_1({\bm r})=0.05n_0$ for ${\bm V}_1=\frac{1}{2}V_{R}\hat{\bm x}$.
Because of the symmetric parameter setting, the second component 
behaves in a similar manner to the first component (not shown) 
[Takeuchi {\it et al.}: Phys. Rev. Lett. {\bf 105}, 205301 (2010), reproduced with permission.
Copyright 2010 by the American Physical Society].}
\label{CSIfig}
\end{figure}
Figure \ref{CSIfig} depicts the characteristic nonlinear dynamics 
of the CSI in an uniform two-component BEC obtained 
by numerically solving the GP equation (\ref{eq:GP}).
In the early stage of the dynamics,
amplification of the unstable modes creates disk-shaped low-density regions
that are orientated in the $x$ direction [Fig. \ref{CSIfig}(b)].
The lowest density inside the disk region reaches zero, creating a local dark soliton.
The soliton in the $i$th component transforms into a vortex ring via snake instability \cite{B.P.Anderson} 
with a momentum antiparallel to the initial velocity ${\bm V}_i$ [Fig. \ref{CSIfig}(c)].
The vortex ring distribution can be determined by the characteristic 
wavenumber of the unstable mode $k_{||} \sim mV_{R}/\hbar$ 
and $k_{\bot} \sim mV_{R}/ 2\hbar$ for a large relative velocity $V_{R}>V_+$.
The length scale $\sim \kappa/V_{R}$ with $\kappa =2\pi \hbar/m$ then 
characterizes both the radii of the vortex rings and the intervals 
between the rings along ${\bm V}_{R}$ immediately after vortex nucleation.
Thus, the vortex line density $l_v$ after the instability is roughly estimated 
to be $l_v \sim V_{R}^2/\kappa^2$, which can be controlled by varying 
the relative velocity $V_{R}$. 

Momentum exchange accelerates after vortex ring nucleation.
The vortex motion then dominates the exchange.
Since the momentum carried by a vortex ring increases with increasing radius,
the radii of the nucleated vortex rings increase with time for momentum exchange.
This dynamics resembles that of quantized vortices under thermal counterflow of liquid helium \cite{Adachi2010}, 
where the vortices are dragged by the mutual friction between the superfluid and normal fluid components.
When the vortex rings become large,
the interaction between the vortex rings deforms the rings and vortex reconnections occur,
which depresses the momentum exchange [Fig. \ref{CSIfig}(d)].
These effects make the vortex dynamics very complicated,
leading to binary QT in which the vortices of both components are tangled with each other [Fig. \ref{CSIfig}(e)].
The momentum exchange almost terminates and each component has an average momentum of nearly zero. 

The relative motion of two-component BECs and the resulting 
CSI can be experimentally realized by employing the Zeeman shift 
of atomic hyperfine states. Hamner {\it et al.} \cite{Hamner} initially prepared overlapping 
two-component BECs of $^{87}$Rb in the hyperfine 
states $|F,m_F \rangle=|1, 1 \rangle$ and $|2, 2 \rangle$,
which satisfy the miscible condition $g_{11}g_{22}>g_{12}^2$. 
When a magnetic field gradient was applied along the longer axis 
of the trap, the gradient generated forces in opposite directions for the two components 
due to the Zeeman shifts. 
Counter superflow then occurred and its relative velocity was controlled 
by the magnetic field gradient \cite{Hamner,Hoefer}.

\subsubsection{Interface instability}
Next, we consider the interface instability of phase-separated two-component 
BECs. The interaction parameters satisfy the immiscible 
condition $g_{11} g_{22} < g_{12}^2$. 
We assume that components 1 and 2 are phase separated at the 
interface near the $y = 0$ plane, which is sustained by the 
external potential $U_i(y)$. 
The density distributions $n_i(y)$ are also assumed to depend only on $y$
and $n_1 = 0$ for $y > \eta$ and $n_2 = 0$ for $y < \eta$, 
where $y=\eta(x,z,t)$ is the interface position. 
Here, we neglect the interface thickness for simplicity. 
The Lagrangian of Eq. (\ref{Lag}) can then be rewritten as
\begin{equation}
L = \int dx dz \left( \int_{-\infty}^{\eta} dy P_1 + \int_{\eta}^{\infty} dy P_2 \right) - \alpha S,
\end{equation}
where $\alpha$ is the interface tension coefficient $\alpha$ \cite{Barankov,Schae}, which
originates from the excess energy at the interface, and 
$S = \int dx dz [ 1 + \left( \partial \eta/\partial x
\right)^2 + \left( \partial \eta/\partial z \right)^2 ]^{1/2} $ is the interface area.
Taking the functional derivative of the action $\int dt L$ with respect
to $\eta(x, z, t)$ and setting it to zero, we obtain
\begin{equation} \label{Ber}
P_1(y = \eta) - P_2(y = \eta) + \alpha \left( \frac{\partial^2
\eta}{\partial x^2} + \frac{\partial^2 \eta}{\partial z^2} \right) = 0,
\end{equation}
which corresponds to the Bernoulli equation in hydrodynamics.

We consider a stationary state in which the $i$th component flows 
with a velocity ${\bm V}_{i} = V_i \hat{\bm x}$ as 
$\Psi_{i0} = \sqrt{n_i(y)} e^{i (m_i V_i x - \mu_i t  )/\hbar }$, 
which is similar to Eq. (\ref{eq:steady}) except that the density depends on $y$. 
Substituting this into Eq. (\ref{Ber}) with $\eta = 0$ gives
the equilibrium condition for the pressure, $g_{11} n_1(0)^2 / 2 = g_{22} n_2(0)^2 / 2$.

To analytically derive the dispersion relation for the interface wave, 
we assume that the system is approximately incompressible. 
We consider the small phase fluctuation $\Psi_i=\Psi_{i0} e^{i \delta \theta_i}$ and the interface 
mode $\eta = \delta \eta$ as 
\begin{eqnarray} \label{dpsi}
\delta \theta_i =  A_i e^{-(-1)^i k y} \cos(k x - \omega t), \\
\delta \eta = a \sin(k x - \omega t),
\label{eta}
\end{eqnarray}
where $A_i$ and $a$ are infinitesimal parameters.
From the kinematic boundary condition, the interface velocity in
the $y$ direction $(\partial / \partial t + V_i \partial / \partial x)
\eta$ must be equal to $\hbar / (i m_i n_i) \Psi_i^* \partial \Psi_i /
\partial y |_{y = \eta}$, giving
\begin{equation} \label{aArel}
-(-1)^i \frac{\hbar}{m_i} A_i k e^{-(-1)^i k \eta} = (V_i k - \omega) a.
\end{equation}
Substituting Eqs. (\ref{dpsi})--(\ref{aArel}) into Eq.~(\ref{Ber}) and
neglecting second and higher orders of $A_i$ and $a$, we obtain
\begin{equation} \label{omega}
\frac{\rho_1}{k} (\omega - V_1 k)^2 - f_1 n_{{\rm s} 1} =
 -\frac{\rho_2}{k} (\omega - V_2 k)^2 - f_2 n_{{\rm s} 2} + \alpha k^2,
\end{equation}
where $n_{{\rm s} 1} = n_1(\eta - 0)$, $n_{{\rm s} 2} = n_2(\eta + 0)$, 
$\rho_i = m_i n_{{\rm s} i}$, and $f_i = dU_i/dy |_{y=\eta}$.
Equation (\ref{omega}) gives the dispersion relation,
\begin{equation} \label{KHId}
\omega = V_{G} k \pm
 \sqrt{-\frac{\rho_1 \rho_2 V_R^2 k^2}{(\rho_1 + \rho_2)^2} +
 \frac{F k + \alpha k^3}{\rho_1 + \rho_2}},
\end{equation}
where ${\bm V}_{G} = (\rho_1 {\bm V}_1 + \rho_2 {\bm V}_2)/(\rho_1 + \rho_2)$ 
is the center-of-mass velocity and $F = n_{{\rm s} 1} f_1 - n_{{\rm s} 2} f_2$ is
the force due to the gradient of the external potential. 
We note that Eq. (\ref{KHId}) has the same form as the dispersion
relation for an interface wave in classical incompressible and inviscid fluids. 
In fluid dynamics, the gradient of the potential is equivalent to gravity. 

\paragraph{Rayleigh--Taylor instability}
First, we consider the case $V_1 = V_2 = 0$. For $F < 0$, the system 
is always dynamically unstable in the wavenumber range $0<k<\sqrt{|F|/\alpha}$, 
which is known as a Rayleigh--Taylor instability \cite{Sasaki,Gautam}. 
This situation corresponds to a layer of a lighter fluid under a heavier fluid layer
in a classical fluid, where the translation symmetry of the interface is spontaneously broken. 

Sasaki {\it et al.} \cite{Sasaki} proposed a system of two immiscible BECs with different 
hyperfine spins (e.g., $| F, m_F \rangle$ = $| 1, -1 \rangle$ and $| 1, 1 \rangle$ of $^{87}$Rb atoms) 
placed in an external magnetic field gradient $B' \equiv dB/dz$. 
Such condensates experience the potentials $+ \mu_{B} B' z/2$ 
and $- \mu_{B} B' z/2$, where $\mu_{B}$ is the Bohr magneton. 
The force generated by this potential can realize 
$F<0$ so that the two condensates are pushed in opposite direction. 
Numerical simulations of the GP equations reveal that
this gradient modulates the interface so that it grows in a mushroom pattern. 
Vortex rings then nucleate due to atoms near the center flowing upward
and atoms at the periphery of the cap of the mushroom shape flowing downward. 
Gautam and Angom \cite{Gautam} considered a system of a $^{85}$Rb--$^{87}$Rb BEC mixture and 
the Rayleigh--Taylor instability caused by tuning the interspecies interaction 
through a Feshbach resonance. The signature of the instability should appear 
in the damping behavior of the collective shape oscillation. 

\paragraph{Richtmyer--Meshkov instability}
The Richtmyer--Meshkov instability occurs when an interface between 
fluids with different densities is impulsively accelerated (e.g., by the passage of a shock wave). 
For atomic BECs, this instability can be caused by a 
magnetic field gradient pulse $B'(t) \propto \delta (t)$ \cite{Bezett}. 
The nonlinear stage of this evolution is qualitatively similar to that of the Rayleigh--Taylor instability. 
However, the instability dynamics differs considerably from that of classical fluids. 
The main difference originates from the quantum surface tension and capillary waves,
which suppress perturbation growth and droplet detachment
from an elongated perturbation finger. The instability for more general time-dependent 
forces has been discussed in Ref. \cite{Kobyakov}

\paragraph{Kelvin--Helmholtz instability}
Finally, we consider a system with the shear flow $V_1 \neq V_2 \neq 0$ ($V_R \neq 0$). 
The dispersion relation Eq. (\ref{KHId}) implies that, 
for $V_{R}^2 > 2 \sqrt{F\alpha} (\rho_1+\rho_2)/\rho_1\rho_2 \equiv V_{\rm KH}^2$,
the imaginary part ${\rm Im}(\omega)$ becomes nonzero and the shear-flow states are dynamically unstable 
against excitation of the interface modes with $k_-<k<k_+$, as in classical Kelvin--Helmholtz instability, 
where $k_{\pm}=k_0\pm\sqrt{k_0^2-F/\alpha}$ with $k_0= \rho_1\rho_2 V_{R}^2 / 2\alpha (\rho_1+\rho_2)$. 
When $F = 0$, $V_{\rm KH}$ vanishes and the system is always dynamically unstable for $V_R > 0$. 
In addition to dynamic instability, thermodynamic instability can occur due to dissipation 
when $\omega < 0$ \cite{TakeuchiKH}. Here, we restrict ourselves to dynamic
instability in nondissipative systems.

\begin{figure}[ht]
\includegraphics[height=0.15\textheight]{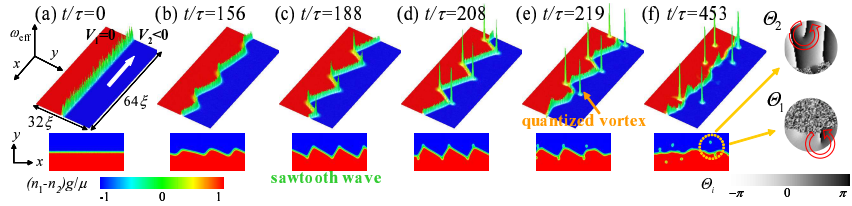}
\caption{Nonlinear dynamics of Kelvin--Helmholtz instability in phase-separated two-component BECs 
with a relative velocity $V_{R}=V_1-V_2>V_{\rm KH}$. 
The numerical simulation of the GP equation was performed with a period $L=64 \xi$
along the $x$-axis and the Neumann boundary condition with $L=32 \xi$ in the $y$-direction, 
where $\xi$ is the healing length. The unit of time is $\xi/c$, where $c$ is the sound velocity.
The initial state (a) is $\Psi^0_i$ with a small random seed to trigger the instability.
(Upper panels) The height and color show the vorticity $\omega_{\rm eff} = (\nabla \times {\bm V}_{G})_z$ 
and the density difference $n_1-n_2$ between the two condensates, respectively.
(Lower panels) Two-dimensional plots of $n_1-n_2$. 
[Takeuchi {\it et al.}: Phys. Rev. B {\bf 81}, 094517 (2010), reproduced with permission.
Copyright 2010 by the American Physical Society]. 
}
\label{KHIfig}
\end{figure}
Figure \ref{KHIfig} demonstrates the Kelvin--Helmholtz instability for $V_{R} >V_{\rm KH}$ \cite{TakeuchiKH}. 
In the linear stage of the instability, the sine wave corresponding to the most unstable mode 
with the maximum imaginary part $\max_k\left\{{\rm Im}\left[\omega(k)\right]\right\}$ is predominantly amplified.
As the amplitude increases, the sine wave is distorted by nonlinearity [Fig. \ref{KHIfig}(b)],
and deforms into a sawtooth wave [Fig. \ref{KHIfig}(c)]. The vorticity $\omega_{\rm eff}$ 
increases on the edges of the sawtooth waves and creates singular peaks [Fig. \ref{KHIfig}(d)].
Subsequently, each singular peak is released into each bulk,
becoming a singly quantized vortex [Fig. \ref{KHIfig}(e)].
The release of vortices reduces the vorticity of the vortex sheet and therefore 
reduces the relative velocity across the interface. 
The released vortices drift along the interface and the system never recovers its initial flat interface.
These nonlinear dynamics differ considerably from those in classical KHI,
where the interface wave grows into roll-up patterns.

The above discussion is valid when the interface thickness $\sim \xi \sqrt{g_{12}/g-1}$ 
(for the case $g_{11} = g_{22}=g$) is much smaller than 
the wavelength of the unstable interface mode, typically given by $\sim \hbar/ m V_R$. 
In the opposite case, the CSI becomes the dominant instability of the flowing state. The crossover 
relative velocity between the two instabilities is evaluated as $V_{R}^c \sim \hbar \sqrt{g_{12}/g-1} / m \xi$ \cite{Suzuki}. 

\section{Conclusions}
Since the realization of BEC in a dilute atomic gas in 1995, most studies on its QHD have been limited to vortex lattices under rotation or motion of a few vortices.
However, as research on superfluid helium has shown, there are many other interesting problems on QHD in atomic BECs, some of which have been discussed in this paper.
Important topics are quantum hydrodynamic instability and QT beyond this instability.
Unlike classical fluid dynamics, most QHD phenomena can be reduced to the motion of quantized vortices.
For example, for QT in atomic BECs, the cascade process of quantized vortices, which transfer energy from large to small scales, can be visualized.
The observation of Kolmogorov spectra could confirm the cascade process in wavenumber space.
The observation of QT in atomic BECs enables us to combine cascade processes in real and wavenumber spaces.
Such investigations are almost impossible in superfluid helium and in classical turbulence.
We anticipate that this research field will develop rapidly in the near future.

\printindex
\end{document}